\documentclass[twocolumn]{svjour3}          
\smartqed  
\usepackage{graphicx}
\usepackage{algorithmic}
\usepackage{amsmath}
\usepackage{amssymb}
\usepackage{amsfonts}
\usepackage{color}
\usepackage[colorlinks=true,
            linkcolor=red,
            urlcolor=blue,
            citecolor=blue]{hyperref}
\usepackage[round]{natbib}
\usepackage{algorithm} 
\usepackage[all]{xy}
\usepackage{color}
\usepackage{DEF-bold}
\usepackage{DEF-script}

\def\pa{{\rm pa}}
\def\pax{{\rm pa}}
\def\ch{{\rm ch}}
\def\tmin{t_{\rm min}}
\def\tmax{t_{\rm max}}
\def\rh{\varrho}

\def\W{{W}}
\def\V{{V}}

\def\XX{\mathbf{X}}
\def\xx{\mathbf{x}}

\def\Pr{\mathbb{P}}

\def\Ind{\mathbb{I}}

\def\p{^\prime}
\def\pp{^{\prime\prime}}
\def\d{{\rm d}}

\begin{document}

\title{Metropolis-type algorithms for Continuous Time Bayesian Networks
}


\author{B{\l}a{\.z}ej Miasojedow \and Wojciech Niemiro \and
        John Noble \and Krzysztof Opalski}

\authorrunning{Miasojedow et. al.} 

\institute{B. Miasojedow \at
Institute of Applied Mathematics, University of Warsaw\\ 
Banacha 2, 02-097 Warsaw,  Poland\\
\email{bmia@mimuw.edu.pl}\and
W. Niemiro \at
Institute of Applied Mathematics, University of Warsaw\\ 
Banacha 2, 02-097 Warsaw,  Poland\\
\email{wniem@mimuw.edu.pl}\and
J. Noble
\at
Institute of Applied Mathematics, University of Warsaw\\ 
Banacha 2, 02-097 Warsaw,  Poland\\
\email{noble@mimuw.edu.pl}\and
K. Opalski\at
Institute of Applied Mathematics, University of Warsaw\\ 
Banacha 2, 02-097 Warsaw,  Poland\\
\email{krzysztof.opalski@mimuw.edu.pl}
}

\date{Received: date / Accepted: date}

\maketitle

\begin{abstract}
 We present a Metropolis-Hastings Markov chain Monte Carlo  (MCMC) algorithm for detecting hidden variables in a {\em continuous time Bayesian network} (CTBN), which uses reversible jumps in the sense defined by ~\cite{green95}. In common with several Monte Carlo algorithms, one of the most recent and important by~\cite{RaoTeh2013a}, our algorithm exploits {\em uniformization} techniques under which a continuous time Markov process can be represented as a {\em marked Poisson process}.  We exploit this in a novel way. We show that our MCMC algorithm can be more efficient than those of likelihood weighting type, as in~\cite{Nod2} and~\cite{FXS} and that our algorithm broadens the class of important examples that can be treated effectively. 
\keywords{Continuous time Bayesian networks \and Markov chain Monte Carlo \and Metropolis algorithms \and uniformization}
\end{abstract}
\section{Introduction}

Continuous time Bayesian networks (CTBNs) represent explicitly 
temporal dynamics in probabilistic reasoning. They were introduced by  
~\cite{Sch} under the name {\em Composable Markov Chains} and then reintroduced by~\cite{Nod1} as {\em Continuous Time Bayesian Networks}.  A {\em continuous time Bayesian network} (CTBN) is a time homogeneous Markov process, which is decomposed into processes whose transition intensities depend on the other processes in the network. The dependence structure between the intensities is encoded by a graph.  CTBNs are only loosely connected with Bayesian networks (BNs); the common feature is that of {\em modularity}, splitting a large problem into smaller components, where the relations between the components are represented graphically. The features represented by the respective graph structures are essentially different, although some of the techniques share similarities. For example, ideas from the {\em intervention calculus} of~\cite{pearl} for BNs transfer quite naturally to CTBN analysis. CTBNs provides a promising and flexible class of models with applications in, for example, Survival Analysis.

While it is relatively straightforward to propose a CTBN as a statistical model, learning algorithms are computationally expensive. There are three categories of learning problem. The first is {\em answering queries}; that is, inserting evidence (information on the trajectory) for some nodes of the CTBN and making probabilistic inference about the behaviour of others when the  structure and parameters are known. This is a classical problem in hidden Markov models (HMMs). The second problem is to estimate the parameters of a 
CTBN, that is the (conditional) transition intensities, based on a {\em learning sample}. The third and by far the most difficult problem is to learn the structure, that is the graph of a CTBN. 

These tasks are `hierarchical' in nature; approaches to the second and third tasks are based on and develop approaches to the first task. Even the first of the these tasks is challenging.   
There are deterministic 
algorithms, for example those proposed by the originators of the model, 
~\cite{Nod1} and developed further by~\cite{Nod4}. The performance of these 
algorithms has been examined through simulated
case studies, but theoretical bounds on the accuracy of  
approximations have yet to be derived. 
~\cite{WNCTBN} presents a {\em randomized} approximation scheme, which uses a deterministic algorithm for static BNs as a subroutine.

Remarkable progress has been achieved via Monte Carlo
(MC) algorithms for CTBNs.~\cite{Nod2} 
 proposed an algorithm which is
a modification of the classical \textit{likelihood weighting} (LW) method 
(\cite{Dagum19971}). Recent results on this method are in~\cite{FXS}. The well-documented drawback of LW is the degeneracy of weights; the likelihood weights tend to be extremely unequally distributed, especially when there is a lot of data available. One remedy is to use  \textit{Markov chain Monte Carlo} (MCMC) techniques
instead of independent sampling, while another is to use sequential Monte Carlo (SMC). An application of this method to CTBNs is given in~\cite{Ng}.

We consider MCMC algorithms based on the Metro\-polis--Hastings scheme.
The idea originated in~\cite{metropolis} and was extended  in ~\cite{HASTINGS} and is at the core of almost all later MCMC developments. Our algorithm makes
 essential use of the representation of sample paths of a continuous time Markov 
process by a marked Poisson process (MPP). This representation, also known under the name of `uniformization', has been exploited 
in many papers; one of the most recent is the work of~\cite{RaoTeh2013a}. 
We use uniformization to construct a variety of proposals including {\rm reversible jumps}, moves which change dimension of the space. 

\section{The basic algorithm}\label{sec: basic}

In this section we describe the basic algorithm in the general setting of
continuous time Markov processes. In the following sections
we use this algorithm as the main building block for more complicated procedures,
specifically designed for CTBNs.

\subsection{Continuous time Markov processes}\label{sub: HMP}

We first recall some definitions and introduce notations which
will be used throughout.
Consider a continuous time stochastic process $(X(t),t\geq 0)$ 
defined on a probability 
space $(\Omega,\F,\Pr)$ with a discrete state space $\X$.
Generic states will be denoted by 
$x,x\p,x\pp,\ldots\in\X$.
Assume that the process is a continuous time, \textit{time homogeneous  Markov process} 
(THMP) with transition probabilities
\begin{equation*}
    P^t(x,x\p)=\Pr\left[X(t+s)=x\p|X(s)=x\right].
\end{equation*}
The initial distribution is denoted by $\nu(x)=\Pr\left[X(0)=x\right]$.
As $\X$ is discrete, $\nu$ can be viewed as a vector and $P^t$  as a matrix (both possibly infinite arrays). 
The \emph{intensity matrix} is 
defined as
\begin{equation*}
    Q(x,x\p)=\lim_{t\to 0} \frac{1}{t} \left[P^t(x,x\p)-I(x,x\p)\right],
\end{equation*}
where $I=P^0$ is the identity matrix.  
Thus $Q=\frac{\d}{\d t} P^t|_{t=0}$.

Expressed  differently, $Q(x,x\p)$ is the intensity of jumps from $x$ to $x\p$: 
\begin{equation*}
    \begin{split}
        \Pr\left[X(t+\d t)=x\p|X(t)=x\right]&=Q(x,x\p)\;\d t\; 
                                                 \text{ for $x\not=x\p$};\\
        \Pr\left[X(t+\d t)=x|X(t)=x\right]&=1-Q(x)\;\d t,
     \end{split}
\end{equation*}
where 
\[Q(x)=-Q(x,x)=\sum_{x\p\not=x}Q(x,x\p)\] denotes the total intensity of  jumping from 
$x$. Clearly, $\sum_{x\p} Q(x,x\p)=0$. The uniformization technique requires that $\sup_x Q(x)$ is bounded. We will assume that this assumption is satisfied.

Equivalently, a THMP satisfying this assumption can be described as a \textit{marked Poisson process}.  
Consider the following sampling algorithm which uses {\em marking} and {\em thinning}. 
Let $\lambda\geq \max_x Q(x)$. At the first stage we sample
\emph{potential} moments of jumps, say $T_1<\cdots<T_i<\cdots$.  These are points
of a time homogeneous Poisson process with intensity $\lambda$. At the second stage, we 
 mark them. The marks, denoted $X_1,\ldots,X_i,\ldots$, are consecutive states of 
the {\em redundant skeleton Markov chain} with transition probabilities

\begin{multline}\label{SkeletonTransitions}  
    P(x,x\p)=\Pr[X_{i}=x\p|X_{i-1}=x]\\=
        \begin{cases}
              Q(x,x\p)/\lambda  & \text{if $x\not=x\p$;} \\
              1- Q(x)/\lambda & \text{if $x=x\p$}.
        \end{cases}  
\end{multline}

\noindent Now let $X(t)=X_{i-1}$ for $T_{i-1}\leq t< T_{i}$ ($i=1,2,\ldots$, with $T_0=0$ and $X_0\sim \nu$). 
The process $(X(t),t\geq 0)$ is a THMP with intensity matrix $Q$ and initial distribution $\nu$.  
In the sequel we fix a finite time interval, say $[\tmin,\tmax]$. In this section, to simplify
notation we set $\tmin=0$ and $\tmax=1$.
The process \[\XXi=(X(t),\;0\leq t \leq 1)\]is thus represented by

\begin{equation}\nonumber
 \XX=\begin{pmatrix} 0 & T_1 & \cdots & T_i & \cdots & T_N & 1\\ X_0 & X_1 & \cdots & X_i & \cdots & X_N & \end{pmatrix},
\end{equation}

\noindent where $N=\max\{n:T_n< 1\}$, with the corresponding  sample path 

\begin{equation}\label{RedundantRepresentation}
 \xx=\begin{pmatrix} 0 & t_1 & \cdots & t_i &\cdots & t_n & 1\\ x_0 & x_1 & \cdots & x_i & \cdots & x_n & \end{pmatrix}.
\end{equation}

\noindent The construction described above is also known under the name {\em uniformization of a THMP}, for example~\cite{hobolth2009} and~\cite{RaoTeh2013a}. The representation \eqref{RedundantRepresentation} 
is {\em redundant} in the sense that there are infinitely many different  double sequences  $\xx$ which 
correspond to the same sample path $\xxi=(x(t),0\leq t\leq 1)$ of the THMP.
We distinguish between an actual sample path $\xxi$ and a representation
$\xx$, by using Greek and Latin letters, respectively. 
The chief advantage of uniformization is that the sequences 
$T_1,\ldots,T_i,\ldots$ and $X_1,\ldots,X_i,\ldots$ are
\textit{independent} of each other. In this section we will work with 
representation \eqref{RedundantRepresentation}.
Let us write the probability density of $\XX$ in the following way:

\begin{multline}\label{PathProbability}
\pi(\xx)\propto \lambda^{n}\d t_1 \cdots \d t_n \Ind(0<t_1<\cdots <t_n<1)\\ 
               \nu(x_0)P(x_0,x_1)\cdots P(x_{n-1},x_{n}).
\end{multline}
This is a slight abuse of notation, since by $\pi(\xx)$ we really mean 
\begin{multline*}
\Pr\big[t_1\leq T_1<t_1+\d t_1, \ldots, t_n\leq T_n<t_n+\d t_n, \\ 
                X_0=x_0,X_1=x_1,\ldots,X_n=x_n\big].
\end{multline*}

\subsection{Hidden Markov models}

Let $(X(t),0\leq t\leq 1)$ be a THMP.
Suppose that process $X(t)$ cannot be observed directly, but that we  observe \textit{evidence} $y$, which is a realisation of a random variable $Y$ with probability distribution $L(.|\xx)$. The quantity $L(y|\xx)$ is  the \textit{likelihood} of evidence $y$ given a trajectory $\xx$.  We assume  that the likelihood only depends
on $\xx$ through the actual sample path  $\xxi=(x(t),0\leq t\leq 1)$; it does not depend on the  `redundant representation'. In this section the concrete form of the evidence (e.g.\ the space
in which $Y$ takes values etc.) is irrelevant.

The problem is to estimate the hidden trajectory $(x(t),$  $0\leq t\leq 1)$ given $y$.  From the
Bayesian perspective, the goal is to compute/approximate the posterior:  
\begin{equation}\nonumber
 \pi(\xx|y)\propto \pi(\xx)L(y|\xx).
\end{equation}
The function $L$, the transition probabilities $Q$ and the initial distribution $\nu$ are assumed 
to be known.

In principle, the problem can be solved by the method of \textit{likelihood weighting} 
(LW), but an inherent problem of LW is the degeneracy of the weights. Even 
in a relatively easy example such as that presented in Section~\ref{sec: simul}, the efficiency of LW is poor. 

\subsection{The Markov chain Monte Carlo algorithm}

We propose a version of the Metropolis-Hastings algorithm  (MHA) which converges to the 
target distribution $\pi(\xx|y)$. The general scheme is standard. At each step of the algorithm, we proceed as follows.
Suppose that $\XX_{m-1}=\xx$. We first sample a \textit{proposal}
$\XX\p=\xx\p\sim q(\xx,\cdot)$. Then let $\XX_{m}:=\xx\p$ (accept the move from $\xx$ to $\xx\p$) with
probablility $a(\xx,\xx\p)$ or let  $\XX_{m}:=\xx$  (reject the move from $\xx$ to $\xx\p$) with probablity 
$1-a(\xx,\xx\p)$. The general MHA recipe for the acceptance probability is
\begin{equation}\label{eqaccept}
    a(\xx,\xx\p)=\min\left(\dfrac{\pi(\xx\p)L(y|\xx\p)q(\xx\p,\xx)}{\pi(\xx)L(y|\xx)q(\xx,\xx\p)},1\right) .
\end{equation}
\floatname{algorithm}{function}
\begin{algorithm}
 \caption{\emph{StepMH$(\xx)$}}

\begin{algorithmic}
    \STATE Sample $\xx\p \sim q(\xx,\cdot)$ {\color{blue}\COMMENT{ proposal }}
    \STATE Sample $U \sim Unif(0,1)$ 
    \IF {$U<a(\xx,\xx\p)$} 
     \STATE {\bf return } $\xx\p$ {\color{blue}\COMMENT{ move accepted with probability $a(x,y)$ }} 
     \ELSE \STATE {\bf return } $\xx$ {\color{blue}\COMMENT{ move rejected with probability $1-a(x,y)$ }} \ENDIF 
    \end{algorithmic}
\end{algorithm}

\noindent Representation \eqref{RedundantRepresentation} suggests several choices of proposal $q$. We describe some of them below.

\subsection*{Change of time}
Let $\xx$ be given by \eqref{RedundantRepresentation}. 
Leaving $n$ and $(x_i)_{i=0}^n$ unchanged, we sample $(t_i\p)_{i=1}^n$  as follows: for a given pair of times $(t_i, t_{i+2})$, we replace $t_{i+1}$ with $t_{i+1}^\prime \sim \mbox{Unif}(t_i, t_{i+2})$, or we can similarly replace several (or all) of the times. The pseudo-code below gives this more precisely. We adopt the convention that $t_0=0$ and $t_{n+1}=1$.


\begin{algorithm}
\caption{\emph{ChangeTime$(\xx)$}}
\label{fun: ChangeTime}
\begin{algorithmic}
    \IF {$n>0$}
    \STATE  Choose a pair $(i_1,i_2)$ from the set $\{0,\ldots,n+1\}$ in such a way that $i_1+1\leq i_2-1$.
    \STATE  Let $t_i\p:=t_i$  for $i\leq i_1$ and for  $i\geq i_2$ 
    \STATE  Sample new times $t_{i_1+1}\p,\ldots,t_{i_2-1}\p\sim Unif(t_{i_1},t_{i_2})$
    \STATE Sort $t_i\p$s  
    \ENDIF  
    \STATE {\bf return } $\xx\p$
    \end{algorithmic}
\end{algorithm}
When this move is proposed, the acceptance probability ~\eqref{eqaccept} reduces to: 

\begin{equation}\label{LikelihoodAcceptance}
    a(\xx,\xx\p)=\min\left(\dfrac{L(y|\xx\p)}{L(y|\xx)},1\right) .
\end{equation}

\noindent This follows from~\eqref{eqaccept} by noting, from~\eqref{PathProbability}, that $\pi(\xx) = \pi(\xx^\prime)$. The points $t_{i_1+1},\ldots,t_{i_2-1}$ are distributed as
a sorted sample from $\mbox{Unif}(t_{i_1},t_{i_2})$. This follows from elementary and well-known properties of the Poisson process. The moves $\xx \rightarrow \xx^\prime$ and $\xx^\prime \rightarrow \xx$, given the choice $(i_1,i_2)$, both amount to choosing uniformly distributed sets of random times with the same number of elements over the same interval. It therefore follows that $q(\xx,\xx^\prime) = q(\xx^\prime, \xx)$ and hence  \emph{ChangeTime} is  $\pi$-reversible; $\pi(\xx)q(\xx,\xx\p)  = \pi(\xx\p) q(\xx\p,\xx)$.

Note that no restriction has been place, so far, on the way that the pair $(i_1,i_2)$ is seleted. There are several possibilities. One way to implement \emph{ChangeTime} is to impose the constraint  $i_1+1 =i_2-1$. This amounts to sampling a single point $t_i\p$. Another extreme
is to choose $i_1=0$, $i_2=n+1$ and sample all $t_i\p$s. The move then loses its `local' character, but is still $\pi$-reversible. 

\subsection*{Change of skeleton}
Let $\xx$ be given by \eqref{RedundantRepresentation}. We leave $n$ and $(t_i)_{i=1}^n$ unchanged and sample $(x_i\p)_{i=1}^n$.
As with `Change of time', there are several possibilities, ranging from `local' to `increasingly global'. Below we describe one  of these; the function \emph{ChangeState} updates
only one state of the skeleton and is actually a variant of the Gibbs Sampler for discrete time chains. 

\begin{algorithm}
\caption{\emph{ChangeState$(\xx)$}}

\begin{algorithmic}
    \IF {$n>0$}
    \STATE  Choose $i$ from the set $\{0,1,\ldots,n\}$
    \STATE  Let $x_j\p:=x_j$  for $j\not= i$
    \IF {$i=0$}
    \STATE Sample new state $x_0\p$ with probability proportional to $\nu(x_0\p)P(x_{0}\p,x_1)$ 
    \ELSIF {$i=n$}    
    \STATE Sample new state $x_n\p$ with probability proportional to $P(x_{n-1},x_n\p)$
    \ELSE
    \STATE   {\color{blue}\COMMENT{ if $0<i<n$ }}  
    \STATE  Sample new state $x_i\p$ with probability proportional to $P(x_{i-1},x_i\p)P(x_i\p,x_{i+1})$
    \ENDIF  
    \ENDIF
    \STATE {\bf return } $\xx\p$
\end{algorithmic}
\end{algorithm}

The value of $i$ is chosen uniformly over $\{0,1, \ldots, n\}$. The {\em proposal} transition clearly defines a $\pi$-reversible chain (by construction) and the acceptance probability is given by~\eqref{LikelihoodAcceptance}.  

If sampling from the Markov bridge is not feasible, then the sampling may be replaced by a Metropolis step which targets this distribution. This can happen, for example, when the  state space is large.

\subsection*{Change of dimension}

We now  proceed to proposals which change the number of marked Poisson points $(t_i,x_i)$. We describe several alternative moves of this type. Their relative merits probably depend on the parameters of the process $Q$, $\nu$ and $L$. We restrict ourselves to the moves which increase / decrease the number of points by one; that is, to algorithms of the following general form:

\begin{algorithm}
\begin{algorithmic}
    \STATE  Sample $d\in \{-1,1\}$ u.a.r.
    \IF  {$d=1$} \STATE Attempt a move that adds one point 
    \ELSIF {$n>0$}  \STATE Attempt a move that erases one point 
    \ELSE \STATE Do nothing   
    \ENDIF
\end{algorithmic} 
\end{algorithm}

Below we describe pairs of moves: a move which creates a new point together with its counterpart which anihilates one point. To clarify the proofs of reversibility, $\xx$ will always denote a configuration of $n$ points given by~\eqref{RedundantRepresentation}, counting neither $t_0=0$ nor $t_{n+1}=1$, while $\xx\p$ will be a configuration of $n-1$ points. When $x_i$ is deleted from $\xx = (x_1, \ldots, x_n)$ to obtain $\xx\p = (x_1\p, \ldots, x_{n-1}\p)$ or vice versa, then $\xx\p$ in terms of the elements of $\xx$, $\xx\p$ is:  $(x_1, \ldots, x_{i-1}, x_{i+1}, \ldots, x_n)$ and this notation will be used.

The proofs that the pairs of moves are reversible will be sketched; the reader is referred to~\cite{green95} for the additional analysis required for full proofs.  

The moves in the following pair: \emph{EraseRandomPoint} and  \emph{AddRandomPoint} are designed to act locally. Suppose a time $t_* \in (t_i,t_{i+1})$ is proposed. We would like the corresponding site $x_* \in {\cal X}$ to be sampled according to the mechanism
\[ \mathbb{P}(x_*) = \frac{P(x_i,x_*)P(x_*,x_{i+1})}{\sum_{y \in {\cal X}} P(x_i,y)P(y,x_{i+1})}.\]

\noindent This would be the natural `Markov bridge' between the two existing points. The denominator, though, may be computationally expensive for large ${\cal X}$ and therefore the simpler proposal $\mathbb{P}(x_*) = P(x_i, x_*)$ is used.

\begin{algorithm}
\caption{\emph{EraseRandomPoint$(\xx)$}}
\begin{algorithmic}
    \STATE Select a subscript  $i\in\{1,\ldots,n\}$ uniformly at random\
    {\color{blue}  \COMMENT{ remove $t_i$ and $x_i$ from $\xx$ }} 
    \FOR {$j:=1$ to $i-1$ }
    \STATE $t_j\p:=t_j$; $x_j\p:=x_j$
    \ENDFOR   
    \FOR {$j:=i$ to $n-1$ }
    \STATE $t_j\p:=t_{j+1}$; $x_j\p:=x_{j+1}$
    \ENDFOR    
    \STATE {\bf return } $\xx\p$
\end{algorithmic}
\end{algorithm}
\begin{algorithm}
\caption{\emph{AddRandomPoint$(\xx\p)$}}
\begin{algorithmic}
    \STATE Sample $t_*$ from $Unif(0,1)$ 
    {\color{blue}  \COMMENT{ add $t_*$ to $\xx\p$ }}
    \STATE Find $i$ such that $t_{i-1}\p<t_*<t_i\p$
    \FOR {$j:=1$ to $i-1$ }
    \STATE   Let $t_j:=t_j\p$ and $x_j:=x_j\p$
    \ENDFOR 
    \STATE Sample $x_*$   from $\X$ according to $\mathbb{P}(x_*) = P(x_{i-1}, x_*)$
    \STATE Let $t_i:=t_*$ and $x_i:=x_*$  
    \FOR {$j:=i$ to $n-1$ }
    \STATE $t_{j+1}:=t_j\p$; $x_{j+1}:=x_j\p$ 
    \ENDFOR
    \STATE {\bf return } $\xx$
\end{algorithmic}
\end{algorithm}

\noindent For these functions, the acceptance are as follows.  
\small
\begin{equation}\nonumber 
\begin{split}
 &a(\xx,\xx\p)\\&=\begin{cases} 
         \min\left(\dfrac{n}{\lambda} \cdot\dfrac{L(y|\xx\p)}{L(y|\xx)}\cdot 
         \dfrac{ P(x_{i-1},x_{i+1})}{ P(x_i,x_{i+1})},1\right), & \text { if } i<n;\\
\min\left(\dfrac{n}{\lambda }  \cdot\dfrac{L(y|\xx\p)}{L(y|\xx)} 
 ,1\right)  & \text { if } i=n,\end{cases}
\\
&a(\xx\p,\xx)\\&=\begin{cases} \min\left(\dfrac{\lambda}{n  } \cdot\dfrac{L(y|\xx)}{L(y|\xx\p)}\cdot 
\dfrac{P(x_i,x_{i+1})}{P(x_{i-1},x_{i+1})},1\right) & \text { if } i<n;\\
\min\left(\dfrac{\lambda}{n  }\cdot\dfrac{L(y|\xx)}{L(y|\xx\p)} ,1\right), & \text { if } i=n.\end{cases}
\end{split}
\end{equation}
\normalsize
Different choices of proposal $\mathbb{P}(x_*)$ for {\em AddRandomPoint} will alter the acceptance rates. 

\begin{proof}[Reversibility of Add/Erase Random Point]
The proof is sketched; a complete proof can be constructed quite easily along the lines found in~\cite{green95}. Let $\xx^\prime$ denote the point with $n-1$ states and $\xx$ the point with $n$ states. Consider $\xx^\prime \rightarrow \xx$ via {\em AddRandomPoint} and $\xx \rightarrow \xx^\prime$ via {\em EraseRandomPoint}. The formulae for $\pi(\xx)$ and $\pi(\xx^\prime)$ are respectively: 

\begin{align*}
\pi(\xx)&=\lambda^n\d t_1\cdots\d t_i\cdots \d t_n\nu(x_0)\cdots P(x_{i-1},x_i)\\&\phantom{\lambda^n\d t_1\cdots\d t_i\cdots \d t_n\nu(x_0)}P(x_{i},x_{i+1})\cdots ,\\ 
\pi(\xx\p)&=\lambda^{n-1}\d t_1\p\cdots\d t_{i-1}\p\cdots \d t_{n-1}\p\nu(x_0\p)\cdots \\&\phantom{\lambda^{n-1}\d t_1\p\cdots\d t_{i-1}\p\cdots \d t_{n-1}\p\nu(x_0\p)}P(x_{i-1}\p,x_i\p)\cdots\\  
     &=\lambda^{n-1}\d t_1\cdots\d t_{i-1}t_{i+1}\cdots \d t_{n}
\nu(x_0)\cdots\\&\phantom{\lambda^{n-1}\d t_1\cdots\d t_{i-1}t_{i+1}\cdots \d t_{n}
\nu(x_0)} P(x_{i-1},x_{i+1})\cdots        
\end{align*}

\noindent while the proposals are: 
\begin{equation}\nonumber
\begin{split}
q(\xx,\xx\p)&=\frac{1}{n},  \quad
q(\xx\p,\xx)&=\d t_i P(x_{i-1},x_i).
\end{split}
\end{equation}

\noindent Therefore, for $i < n$, 

\begin{multline}\nonumber
\dfrac{\pi(\xx\p)L(y|\xx\p)q(\xx\p,\xx)}{\pi(\xx)L(y|\xx)q(\xx,\xx\p)}=
  \dfrac{n}{\lambda} \cdot\dfrac{L(y|\xx\p)}{L(y|\xx)}\cdot 
         \dfrac{P(x_{i-1},x_{i+1})}{ P(x_i,x_{i+1})}
\end{multline}
and the conclusion follows; similarly for $i = n$.
\end{proof}

Finally, we propose a pair of moves which is essentially a restricted version of the previous pair, where we only allow a change at a \textit{virtual} jump point. These moves are {\em EraseVirtualPoint} and {\em AddVirtualPoint}. The first of these removes a point $x_i$ from the skeleton if and only if it is a {\em virtual point}, i.e. if and only if $x_{i-1} = x_i$. The second of these adds a virtual point into the skeleton. These may be used instead of {\em EraseRandomPoint/AddRandomPoint}, provided $\lambda$ is chosen so that $1 - \frac{1}{\lambda}\max_x Q(x)$ is sufficiently large; if $\max(1 - \frac{1}{\lambda} Q(x))$ is close to zero, then there will be few virtual points; the {\em EraseVirtualPoint} algorithm will rarely find a virtual point and the {\em AddVirtualPoint} algorithm will reject any proposal with probability close to $1$; these algorithms will neither add or remove virtual points. The choice of $\lambda$ is therefore an interesting problem. In~\cite{RaoTeh2013a}, an appropriate choice of $\lambda$ is made and only virtual points are added or removed. 



\noindent  

\subsection{Piecewise homogeneous processes}\label{sub: piece-wise}

Our applications to CTBNs require a generalisation of the algorithms 
in previous subsections so that they can treat a Markov process which `switches from one regime to another'; the process itself is not time homogeneous, but it is piecewise time homogeneous.
Let $\tmin<r_1<\ldots <r_k<\tmax$ be points which partition the interval $[\tmin,\tmax]$ into
sub-intervals $[r_{j-1},r_j]$, $j=1,\ldots,k,k+1$, with $r_0=\tmin$ and $r_{k+1}=\tmax$.
Assume that $(X(t),\tmin\leq t\leq \tmax)$ is a Markov process with finite state space $\X$, 
such that  $(X(t),r_{j-1}\leq t\leq r_j)$ is a \textit{homogeneous} Markov process with
intensity matrix $Q_j$. The end value of $X$ on $[r_{j-1},r_j]$ is the initial value of $X$ on $[r_{j},r_{j+1}]$;  $X(r_j-) = X(r_j+) = X(r_j)$. Let $\nu$ be the probability distribution
of $X(0)$. Below we describe the {\em redundant representation}, also known as the {\em uniformization} construction of such chains.

Begin with choosing the  {\em redundant intensities of jumps} $\lambda_j\geq \max_x Q_j(x)$, where $Q_j(x)=-Q_j(x,x)$. 
The potential jump times $\tmin<T_1<\cdots<T_i<\cdots<T_n<\tmax$ are then sampled from a piece-wise 
homogeneous Poisson process with intensity $\lambda_j$ on $[r_{j-1},r_j]$.
This can be done using the algorithm below.

\begin{algorithm}
\begin{algorithmic}
\FOR {$j:=1$ to $k+1$ }
 \STATE Sample $n_j$ from $Poisson(\lambda_j (r_{j}-r_{j-1}))$ distribution.
 \FOR {$l=1$ to $n_j$}
 \STATE Sample $T_j^{(l)}$
 from  Uniform$(r_{j-1},r_{j})$.
 \ENDFOR
 \ENDFOR
 \STATE Sort all points $T_j^{(l)}$, $l=1,\ldots,n_j$, $j=1,\ldots, k+1$ in   increasing order and rename them 
   $T_1,\ldots,T_n$.
\end{algorithmic}
\end{algorithm}

\noindent where  $n=\sum_j n_j$. Some of $n_j$s (or even all of them) can be $0$.
At the second stage, we mark  points that have been generated $T_1=t_1,\ldots,T_n=t_n$ by simulating a 
 {\em redundant skeleton Markov chain} $X_1,\ldots,X_n$, similarly to the homogeneous case. 
Here the chain is no longer time homogeneous and its transition probabilities
depend on $(t_i)_{i=1}^n$. More precisely, they depend on the intervals to which subsequent $t_i$s belong.
Write 
\begin{equation}\nonumber
    P_j(x,x\p)=
        \begin{cases}
              Q_j(x,x\p)/\lambda  & \text{if $x\not=x\p$;} \\
              1- Q_j(x)/\lambda & \text{if $x=x\p$}
        \end{cases}    
\end{equation}
and let
\begin{equation}\label{SkeletonNonhomo}
  \Pr[X_{i}=x\p|X_{i-1}=x]=  P_j(x,x\p) 
\end{equation}
whenever $ t_i\in [r_{j-1},r_j)$.

The rest of the construction is exactly the same as for time homogeneous processes. Let $X(t)=X_{i-1}$ for $t_{i-1}\leq t< t_{i}$, ignoring the `change of regime' points 
$r_j$. Nota bene: we tacitly assumed that these points $r_j$ are fixed. In our applications to CTBNs,
the change of regime at a node occurs if some parent node changes state, so the $r_j$s are
random. This does not present any difficulty, since the whole construction may be applied conditionally. 

The proposal moves and accompanying acceptance rules described in the previous subsections
are easy to modify for piecewise homogeneous process. We briefly describe the necessary modifications, omitting the details, which are rather self-evident. In the sequel, write $r(t)=j \quad\text{if } t\in [r_{j-1},r_j)$.

\begin{itemize}
\item \emph{ChangeTime} needs no modifications if it is applied \textit{separately} to any interval of homogeneity 
$[r_{j-1},r_j]$. This means that  $r_{j-1}$ and $r_j$ take over the role of the endpoints $0$ and $1$, respectively and we move points $t_i$ belonging to $(r_{j-1},r_j)$. Thus a jump time $t_i$ never moves from one prior regime to another. 

\item Alternatively, \emph{ChangeTime} can be applied \textit{globally} to the whole interval $[\tmin,\tmax]$ on the 
condition that we use `uniform uniformization' on this interval; that is, we choose $\lambda_j=\lambda\geq \max_j\max Q_j(x)$ then a jump time  $t_i$  can be moved to a different interval of homogeneity.

\item   In \emph{ChangeState} we use the skeleton transition probabilities linked to the jump times instead of a single $P$. More precisely, when updating $x_i$ to $x_i\p$, we sample the new state $x_i\p$ with probability proportional to $P_{r(t_i)}(x_{i-1},x_i\p)P_{r(t_{i+1})}(x_{i}\p,x_{i+1})$ if $0<i<n$. Similarly, $x_0\p$ is sampled with  probability proportional to  $\nu(x_0\p)P_{r(t_1)}(x_{0}\p,x_1)$ and  $x_n\p$ with probability proportional to $P_{r(t_n)}(x_{n-1},x_n\p)$.

 \item In \emph{AddRandomPoint} and \emph{EraseRandomPoint} we modify the acceptance probabilities analogously. 
\end{itemize}

\section{Continuous time Bayesian networks}\label{sec: CTBN}

First we recall the definition of a CTBN and basic facts about this notion.

\subsection{Definitions and notations}\label{sub: defno}
 
Let $(\V,\E)$ be a directed graph with possible cycles.
We write $v\to w$ instead of $(v,w)\in\E$, whenever the graph is fixed. 
For $v\in\V$ let $\pa(v)=\{w:w\to v\}$ be the set of parents of $v$.
Suppose $\A_v$ is the alphabet of possible states of node $v$.
We consider a class of continuous time stochastic processes on the product
space $\X=\prod_{v\in\V} \A_v$. Thus a state $x\in\X$ is a
configuration $x=(x_v)=(x_v)_{v\in\V}$, where $x_v\in\A_v$. As usual,
if $\W\subseteq\V$ then we write $x_\W=(x_v)_{v\in\W}$ for 
the configuration $x$ restricted to nodes in $\W$. We also use the notation $\X_W=\prod_{v\in\W} \A_v$, so that $x_W\in\X_W$. 
The set $\W\setminus\{v\}$ will be denoted by $W-v$ and $\V\setminus\{v\}$ simply by $-v$. 
Suppose we have a family of functions
$Q_v:\X_{\pa(v)}\times(\A_v\times \A_v)\to[0,\infty )$.
For fixed $c\in \X_{\pa(v)}$, we consider $Q_v(c;\cdot,\cdot)$ as a  conditional intensity matrix 
(CIM) at node $v$ (only the off-diagonal elements of this matrix
have to be specified since those on the diagonal are irrelevant).
The state of a CTBN at time $t$ is a random element $X(t)$ of the space 
$\X$ of configurations. Let $X_v(t)$ denote its $v$th coordinate. 
The process ${\XXi}=\left((X_v(t))_{v\in\V},t\geq 0\right)$ 
is assumed to be Markov and its evolution can be described 
informally as follows. The transition intensities at node $v$ depend on the current 
configuration of the parent nodes. If the parent configuration changes, then node $v$ switches to other transition 
intensities.  If  $x_v\not=x_v\p$ then  

\begin{multline*}
         \Pr\left[X_v(t+\d t)=x_v\p|X_{-v}(t)=x_{-v},X_v(t)=x_v\right]=\\
              Q_v(x_{\pa(v)}; x_v,x_v\p)\,\d t. 
\end{multline*}

\noindent Formally, CTBN is a THMP with transition intensities given by  

\begin{multline*}
    Q(x,x\p)=\\
          \begin{cases}
             Q_v(x_{\pa(v)};x_v,x_v\p) & \text{if $x_{-v}=x_{-v}\p$} \\ & \text{and $x_{v}\not=x_{v}\p$ for some $v$;} \\        
              0       &  \text{if $x_{-v}\not=x_{-v}\p$ for all $v$,}
          \end{cases}
\end{multline*}

\noindent for $x\not=x\p$ (of course, $Q(x,x)$ must be defined  by subtraction in the usual way to ensure that $\sum_{x\p} Q(x,x\p)=0$).  
\goodbreak

\subsection{Probability densities of CTBNs}\label{sub: dens}

An important special case of  evidence is the {\em complete} observation of some nodes of the CTBN. To compute the posterior distribution
over unobserved nodes, we need the likelihood; the probability density of the observed trajectories. 
Formulae for densities of general HMMs can be obtained from \eqref{PathProbability} 
by `intergrating out the virtual jumps'. Such formulae appear in many papers, e.g. \cite{Nod2} and \cite{RaoTeh2013a}, equation (2).
The latter reference  also contains a comprehensive discussion about the reference measure with respect to which the density is computed. This is not important for our purposes. Below, we recall a formula specialized to CTBNs, which is equivalent to expressions given in \cite{Nod2}.  
As before, we set a finite time horizon, say $[\tmin,\tmax]=[0,1]$, and consider a CTBN process

\[{\XXi}=\left((X_v(t))_{v\in\V},0\leq t\leq 1\right).\]

\noindent Recall that the state space is $\X=\prod_v \A_v$ and the transition intensities are described by CIMs $Q_v$. We need the following notations: 

\begin{itemize}
  \item[] Let $n_v^{\xxi}(c;\;a,a\p)$ denote the number of jumps from 
$a\in\A_v$ to $a\p\in\A_v$  at node $v$, 
which occurred when the parent configuration was $c\in\X_{\pax(v)}$.
\item[] Let $t_v^{\xxi}(c;\; a)$ be the length of time that node $v$ was in state $a \in \A_v$   and the parent configuration was $c\in\X_{\pax(v)}$.   
\end{itemize}

\noindent The density of the sample path ${\xxi}=\left((x_v(t))_{v\in\V},0\leq t\leq 1\right)$ is the following:

\begin{equation}\label{CTBNdens}
\begin{split}
        p({\xxi})=\nu\left(x(0)\right)
             \prod_{v\in\V} \rh({\xxi}_v\Vert {\xxi}_{\pax(v)}),         
\end{split}
\end{equation}
where $x(0)=(x_v(0))_{v\in\V}$ is the configuration at time 0, $\nu$ is the initial distribution and
\begin{equation}\label{cbi}
\begin{split}
       \rh({\xxi}_v\Vert {\xxi}_{\pax(v)})&=\\
             &\bigg\{\prod_{c\in\X_{\pax(v)}}
                      \prod_{a\in\A_v} \prod_{a\p\in\A_v\atop a\p\not=a} 
                      Q_v(c;\; a,a\p)^{n_v^{{\xxi}}(c;\; a,a\p)}\bigg\}\\
             &\bigg\{\prod_{c\in\X_{\pax(v)}}
                       \prod_{a\in\A_v} \exp\left[-Q_v(c;\; a)
                              t_v^{{\xxi}}(c;\; a)\right]\bigg\},
\end{split}
\end{equation}
To give a clear and useful 
interpretation of \eqref{cbi}, let us recall the notion of
\emph{conditioning by intervention}. This concept is well understood
in the context of static BNs and {acyclic} digraphs  
(\cite{lauritzen} and the references therein). The idea can easily be  carried over to CTBNs with graphs that are possibly cyclic. 
For $\W\subset\V$, let 
${\XXi}_W=\left((X_v(t))_{v\in\W},0\leq t\leq 1\right)$ be the process
restricted to the nodes in $\W$. Write
 $\pax(\V\setminus\W)=
\{w: w\in\W\text{ and }w\to v\text{ for some }v\not\in\W\}$ and 
\begin{multline}\label{cbigen}
\rh({\xxi}_{\V\setminus \W}\Vert{\xxi}_{\W})
=\rh({\xxi}_{\V\setminus \W}\Vert{\xxi}_{\pax(\V\setminus\W)})       
\\= \prod_{v\in\V\setminus\W} \rh({\xxi}_v\Vert {\xxi}_{\pax(v)}).         
\end{multline}
Suppose that
the trajectory $\XXi_{\W}=\xxi_{\W}$ is fixed. Imagine that we remove
the arrows of the graph $(\V,\E)$ which lead into $\W$ and allow the nodes 
outside $\W$ to evolve according to the CTBN dynamics, always using the 
current values of the fixed trajectory  $\xxi_\W$ in the CIMs. Strictly 
speaking, the resulting stochastic process on $\X_{\V\setminus \W}$ is a piecewise time homogeneous Markov chain where the intensity matrix is constant on time intervals where the configuration $x_\W(t)$ is constant. 
If we start at a deterministic initial state, say $x_{\V\setminus \W}(0)$, 
 then the density of the process is proportional to
$\rh({\xxi}_{\V\setminus \W}\Vert{\xxi}_{\pax(\V\setminus\W)})$. 
Thus \eqref{cbigen} corresponds to the condition-by-intervention
transition rule of a CTBN on $\X_{\V\setminus \W}$, given $\xxi_{\W}$.  

\section{A Gibbs sampler for CTBNs}\label{sec: defex}

The main idea is a straightforward application of `Metro\-polis within Gibbs'.  We embed the algorithms from the previous section into a GS which updates single nodes one after another.  In what follows, we focus on a special form of evidence.
Assume that $\W\subset\V$ is the set of  nodes which are observed completely over some period of time, say $[0,1]$.
We update a node $v\not\in\W$  using one or more Metropolis steps which  preserve probability distribution

\begin{equation}\nonumber
 p({\xxi}_{v}|{\xxi}_{-v})\propto p({\xxi}_{v},{\xxi}_{-v})=p(\xxi).
\end{equation}

\noindent This general strategy has many variants.
We can use either a random scan or a systematic scan Gibbs sampler (GS). The choice of
Metropolis moves and the number of such moves for a single Gibbs step can be specified 
in many ways. Instead of single nodes we can update subsets of ${\V\setminus \W}$
simulataneously (block GS).  The effeciency of such variants is clearly  problem-dependent and will not be discussed here. We will only explain how to apply the Metropolis moves from Section~\ref{sec: basic} to GS for CTBNs.

To apply functions from Section  \ref{sec: basic}, we need to specify the prior distribution  at node $v$, which is the distribution of a piecewise homogeneous Markov process with state space $\A_v$, and the likelihood. In Section~\ref{sec: basic}, we worked with redundant 
representations $\xx$ rather than sample paths $\xxi$. This causes no problem, since any move that preserves the probability distribution of an MPP also preserves the inherited distribution of the Markov process represented by the MPP. Let us therefore abuse notation and use Greek letters as arguments of $\pi$ and $L$. With this convention, we   express $p(\xxi)\propto \pi(\xxi_{v})L({\xxi}_{-v}|{\xxi}_{v})$. 
To implement efficient algorithms over a CTBN, we have to assume some 
special stucture of the initial distribution $\nu$.
We sketch two scenarios below. 

In some applications we may assume, as in \cite{Nod1} 
  and \cite{WNCTBN}, 
that the initial distribution is specified as a static
Bayesian network. Suppose that $(\V,\E_0)$ is a \textit{directed acyclic graph} 
(DAG)  
and that the distribution $\nu$ factorizes as:

\begin{equation}\label{asBN}
    \nu\big((x_{v}(0))_{v\in\V}\big)=\prod_v \nu(x_{v}(0)|x_{\pax^0(v)}(0)),
\end{equation}

\noindent where $\pax^0(v)$ refers to the set of parents with respect to $\E_0$. The condition-by-intervention initial distribution is then defined in the standard way:

\begin{equation}\nonumber
\nu\big(x_{\V\setminus\W}(0))\Vert x_{\W}(0)\big)=\prod_{v\not\in\W} \nu(x_{v}(0)|x_{\pax^0(v)}(0)).
\end{equation}

\noindent If \eqref{asBN} holds, we may choose

\begin{equation}\nonumber\label{decompositionBN}
\begin{split}
\pi(\xxi_v)&=p(\xxi_v\Vert \xxi_{-v})= \nu(x_{v}(0)\Vert x_{-v}(0))\rh(\xxi_v\Vert \xxi_{-v}),\\
 L({\xxi}_{-v}|{\xxi}_{v})&=p(\xxi_{-v}\Vert \xxi_{v})=\nu(x_{-v}(0)\Vert x_{v}(0))\rh(\xxi_{-v}\Vert \xxi_{v}).
\end{split}
\end{equation} 
Of course, $\rh(\xxi_v\Vert \xxi_{-v})$ depends on $\xxi_{-v}$ only through $\xxi_{\pa(v)}$ and
 $\rh(\xxi_{-v}\Vert \xxi_{v})$ depends on $\xxi_{-v}$ only through $\xxi_{\ch(v)}$. Similarly,
the $\nu$ terms depend only on the variable / parent configurations $x_v(0) / x_{\pa^0(v)}(0)$ where 
$\pa^0$  denotes the parent set with respect to the DAG $\E_0$. 
The likelihood is computed according to formulae in Section~\ref{sec: CTBN}.
It follows that the functions of Section~\ref{sec: basic} can be implemented efficiently.

Another possible scenario 
is that we are able to evaluate the full conditional distributions at time 0,  $\nu(x_v(0)|x_{-v}(0))$. 
Then we can write:

\begin{equation}\nonumber\label{decompositionFC}
\begin{split}
\pi(\xxi_v)&= \nu(x_{v}(0)|x_{-v}(0))\rh(\xxi_v\Vert\xxi_{-v}),\\
 L({\xxi}_{-v}|{\xxi}_{v})&= \nu(x_{-v}(0))\rh(\xxi_{-v}\Vert\xxi_{v})\propto \rh(\xxi_{-v}\Vert\xxi_{v}).
\end{split}
\end{equation} 

\noindent The factor $ \nu(x_{-v}(0))$ need not be evaluated.

The Metropolis within GS produces a sequence of processes 
${\XXi}_{\V\setminus \W}^1,\ldots,{\XXi}_{\V\setminus \W}^m,\ldots$ which
is a Markov chain with stationary distribution $\pi(\cdot|\xxi_{\W})$.
The estimator of the inference probability is 
\begin{equation}\label{est}
    \hat \pi_m(\cdot|{\xxi}_{\W})=\frac{1}{m}\sum_{j=1}^m 
               \Ind \left\{{\XXi}_{\V\setminus \W}^j\in \cdot\right\}.
\end{equation} 
The funcions of Section \ref{sec: basic} which change the number of jumps ensure that the chain is irreducible and thus
ergodic. Estimator \eqref{est} is strongly consistent.

\section{Simulation results}\label{sec: simul}

We applied  our algorithm to a simple network. Consider a CTBN with two vertices $X$ and $Y$, each 
of them binary, i.e.\ the alphabet of states is $\A_X=\A_Y=\{1,2\}$. The process at node $X$ is hidden, 
node $Y$ is fully observable. The graph is
$$ X \longrightarrow Y$$

The parameters of the process are 
\begin{itemize}
   \item the transition intensity matrix $Q_X(x,x\p)$;
   \item the conditional intensity matrices $Q_Y(x;y,y\p)$, where $x$ denotes the state of the parent node $X$.
\end{itemize}

\noindent We experimented with two sets of parameters. In \ Example 1, the 
current state of $X$ affects the probabilities with which $Y$ `chooses its state'.
In  Example 2, the current state of $X$ affects the frequency of jumps of $Y$.
To give the algorithm a fighting chance to restore the path of $X$ given information
about $Y$, we have to choose transition rates of $Y$ substantially higher than that of $X$. 
In both exapmples we used the LW algorithm and our Metropolis-type algorithm. In the latter,
we iterated functions \emph{ChangeTime}, \emph{ChangeState}, \emph{AddRandomPoint} and \emph{EraseRandomPoint} 
cyclically. 

\subsection*{Examples 1 and 2} Consider a THMP $(X,Y)$ with state space $\{1,2\}^2$, where $X$ itself is a THMP with  transition matrix  
\begin{equation}\nonumber
 Q_X=\begin{pmatrix}  -4  &  4 \\
                       5  & -5 \end{pmatrix}
\end{equation}

\noindent and the transition matrix for $Y$ depends on $X$. $Y$ is observed and we want to make inferences about the hidden Markov chain $X$. For Example 1, we take:  
\begin{equation}\nonumber
 Q_{Y|X=1}=\begin{pmatrix}  -100  &  100 \\
                              20  & -20 \end{pmatrix},\qquad
 Q_{Y|X=2}=\begin{pmatrix}   -20  &  20 \\
                             100  & -100 \end{pmatrix}
\end{equation}

\noindent so that for $Y$ has a substantially greater probability of being in state $2$ when $X = 1$ than when $X = 2$ and has substantially greater probability of being in state $1$ when $X = 2$ than when $X = 1$.  This is illustrated by a sample path of the process $(X,Y)$ shown in Fig.~1 
The time interval under consideration is always 
$[\tmin,\tmax]=[0,1]$.

For Example 2, we take the same $Q_X$, but with $Q_{Y|X}$ given below. 

\begin{equation}\nonumber
 Q_{Y|X=1}=\begin{pmatrix}  -100  &  100 \\
                             100  & -100 \end{pmatrix},\qquad
 Q_{Y|X=2}=\begin{pmatrix}   -2  &  2 \\
                              2  & -2 \end{pmatrix}
\end{equation}

\noindent The jump frequency for $Y$ is substantially higher for  $X=1$  than for  $X=2$, but the jumps 
from $Y=1$ to $Y=2$ and from $Y=2$ to $Y=1$ are equally likely. 
This is illustrated by the sample paths of the process shown in Fig.~2 
The type of information transmitted from $X$ to $Y$ is clearly quite different in these two examples.

\begin{figure}\label{fig: OldXY}
\centering
 \includegraphics[width=8cm]{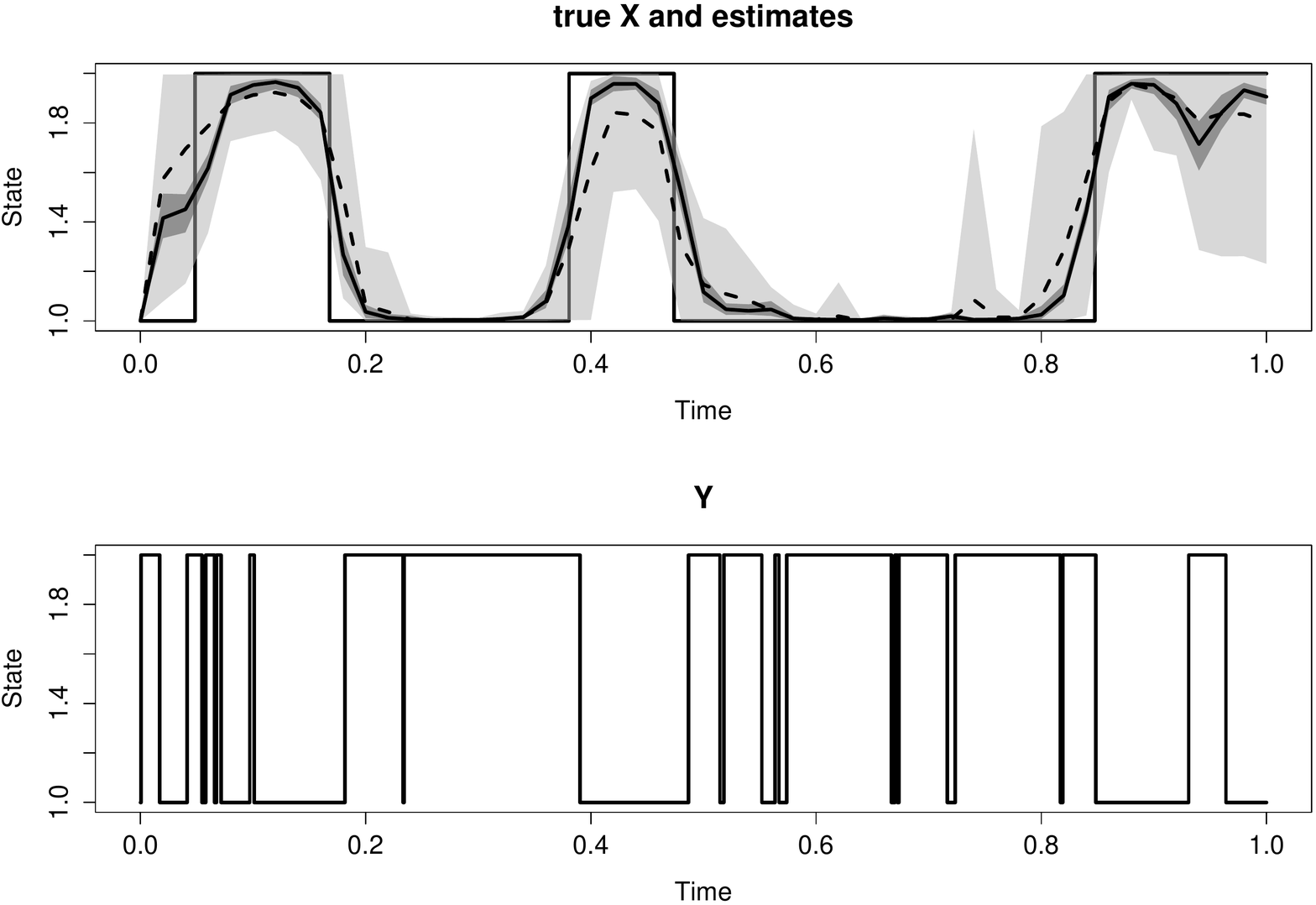}
\caption{Results of Estimation (above), Sample paths of $X$ and $Y$ (below) for Example 1. Above: Dashed - likelihood weighting, solid- MCMC, lighter shadow - variability of likelihood weighting, darker shadow - variability of MCMC}
 \end{figure}
 
\begin{figure}\label{fig: NewXY}
\centering
 \includegraphics[width=8cm]{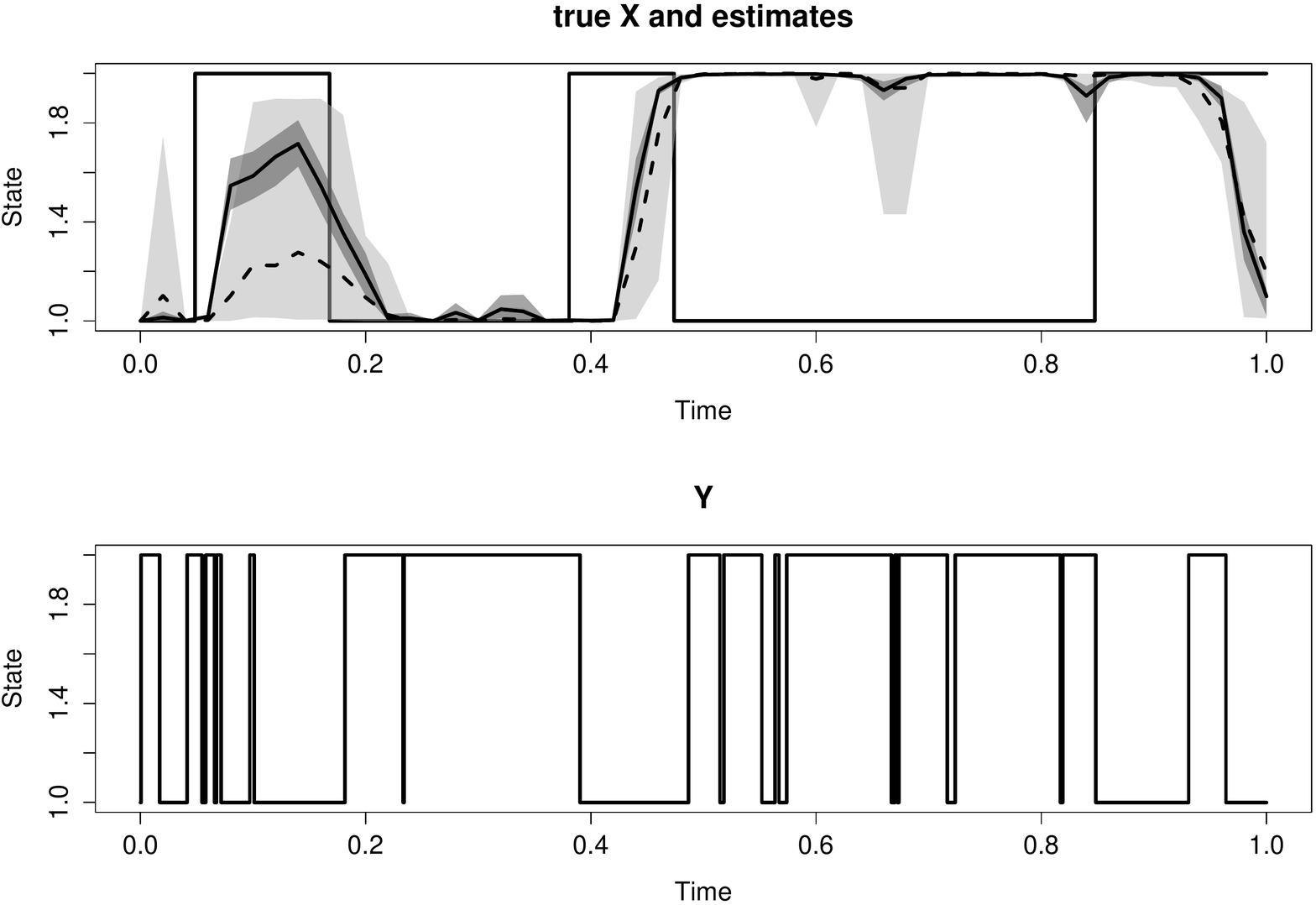}
\caption{Results of Estimation (above), Sample paths of $X$ and $Y$ (below) for Example 2. Above: Dashed - likelihood weighting, solid- MCMC, lighter shadow - variability of likelihood weighting, darker shadow - variability of MCMC}
 \end{figure}

Fig.~1 
shows the estimation results for Example 1, while Fig.~2 
shows the estimation results for Example 2. More precisely, these figures depict the posterior probability of $X(t)=x$ given the whole path $(Y(t), 0\leq t \leq 1)$.
The results obtained by LW are given by dotted lines, while the results of our MCMC algorithm are given by dashed lines. 
The LW estimator and our MCMC algorithm given by~\eqref{est} are applied to $\hat \pi_m(X(t)=1|\YY=\eeta)$
for a grid of $t$-points, where $\YY=\eeta$ is the observed path of $Y$.  
The solid broken line is the true unobserved path of the hidden node $X$.


 This experiment gives a spectacular illustration of the degeneracy of weights  phenomenon of the LW algorithm.
For a sample of size $m=10000$,
the cumulative sums of the 10 largest weights (normalized to sum to 1) are shown in Table~1 
for Example 1 and in Table~2 
for Example 2.
\begin{table}[!ht]
\label{tab:1}

\begin{tabular}{|c|c|c|c|c|}
\hline
 0.538 & 0.906 & 0.939 & 0.955 & 0.967 \\\hline 0.974 & 0.981 & 0.984 & 0.986 & 0.988\\
\hline
\end{tabular}
\caption{Cumulative sums of the 10 largest weights in the LW algorithm for Example 1}
\end{table}

\begin{table}[!ht]\label{tab:2}
\begin{tabular}{|c|c|c|c|c|}
\hline
  0.589 & 0.741 & 0.781 & 0.803 & 0.825 \\\hline0.847 & 0.867 & 0.886 & 0.899 & 0.912\\
\hline
\end{tabular}
\caption{Cumulative sums of the 10 largest weights in the LW algorithm for Example 2}
\end{table}

\noindent Thus, for Example 1, 10 of 10000 points carry about 98.8\% of the total mass. Roughly speaking, $9990$ sampled points are effectively useless. The results for Example 2 are slightly better; the l0 largest weights account for 91.2\% of the value. In a more realistic situation, when the problem is to compute the posterior over a set of hidden nodes of a complicated CTBN, the phenomenon of degeneracy may become substantially worse.

For both Example 1 and Example 2, the number of steps of the Metropolis-type algorithm is $m=10000$. 
Each step consists of consecutive applications of 
\emph{ChangeTime}, \emph{ChangeState} and then \emph{AddRandomPoint} or \emph{Erase\\RandomPoint}. 
Parameter $\lambda$ is chosen as $2.5 \times \max_x Q_X(x)$. 
The acceptance rate was approximately $0.5$ in the experiment for both examples. The results were similar for both examples, even though the type of evidence was quite different.  
The results represent a substantial improvement over those for LW.



 
\subsection*{Example 3: A Lotka Volterra Model} Let $(X,Y)$ be a Markov chain with intensities: 
\begin{align*}
&Q_{(x,y) \rightarrow (x,y+1)} = y(\alpha - \beta y) \vee 0\;,\\
&Q_{(x,y) \rightarrow (x,y-1)}= \gamma yx\wedge M\;,\\
&Q_{(x,y) \rightarrow (x,z)} = 0\ \text{for $z \neq y\pm 1$, }\\
&Q_{(x,y) \rightarrow (x+1,y)} =  \delta x y \wedge M\;,\\
&Q_{(x,y) \rightarrow (x-1,y)} = \eta x \wedge M\;,\\
&Q_{(x,y) \rightarrow (z,y)} = 0\ \text{ for $z \neq x \pm 1$,}
 \end{align*}
where $\alpha,\beta,\gamma,\delta, \eta$ are non-negative constants and $M$ is a truncation parameter so that uniformization may be used. In this model, $Y$ is the {\em prey}, while $X$ is the {\em predator}. The prey is {\em hidden}, while the predator is {\em observed}. In the absence of predation, the prey population grows, with limitations on the growth rate due to the carrying capacity of the environment. At the same time, each prey is killed off with intensity proportional to the number of predators. In the absence of any prey, the predator's death rate is exponential, while the prey contributes to the predator's growth rate. The coefficients are chosen so that the Markov chain $(X,Y)$ has a stationary distribution.  Such models are discussed in~\cite{murray}. The results are illustrated in Fig.~3. 

This example illustrates that, although the intensities need to remain bounded so that uniformization may be used, we do not need a bounded state space.

\begin{figure}\label{figLotVolt}
\centering
 \includegraphics[width=8cm]{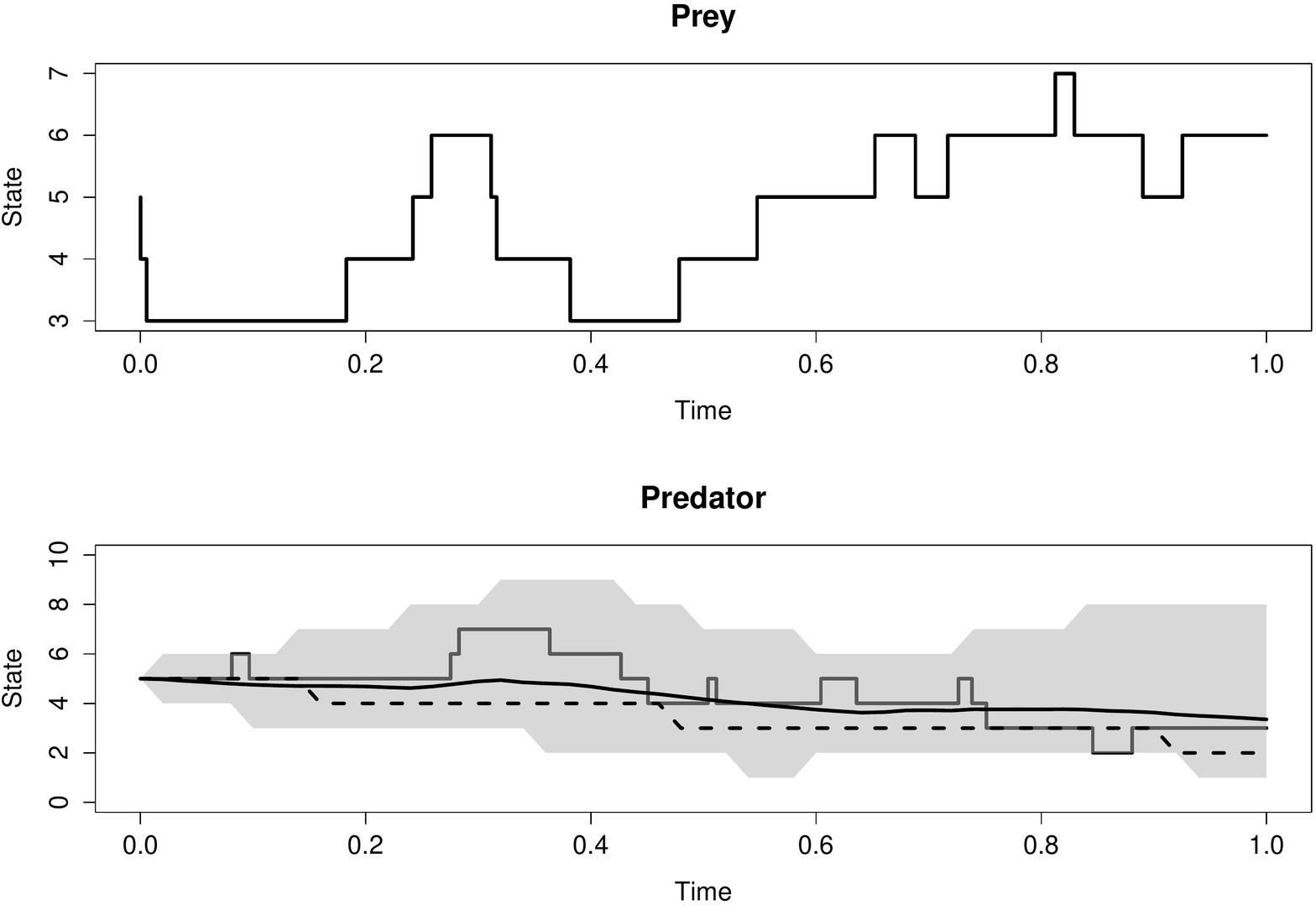}
\caption{Results for Lotka Volterra example: dashed line - posterior median, solid - posterior mean, shadow - 0.1 and 0.9 posterior quantiles}
 \end{figure}
 
\section{Discussion} We have presented a new Metropolis-Hastings MCMC algorithm for detecting hidden variables in a CTBN. The algorithm presented here has some similarities to that of~\cite{RaoTeh2013a}, but operates on essentially different principles. The two algorithms are suited to different situations. Our algorithm is substantially more local in flavour. Firstly, in contrast to Rao and Teh, each move of {\em Add/EraseRandomPoint} in our algorithm only needs to update the sufficient statistics for three points, while a single move of the Rao and Teh algorithm re-evaluates the entire trajectory. The Rao-Teh algorithm uses an FFBS approach which requires, at each step the multiplication of transition matrices, with a cost of ${\cal O}(n|{\cal X}|)$ where $|{\cal X}|$ is the number of elements in the state space and $n$ is the number of jumps, including {\em virtual} jumps. Broadly speaking, the Rao-Teh algorithm outperforms our algorithm in situations with a state space of moderate size, but the cost of the Rao - Teh algorithm increases linearly with the size of the state space and it  cannot perform reliably with {\em infinite}  state space; it is well known that the stationary distribution for a truncated problem may be substantially different from the target stationary distribution; the Rao-Teh algorithm may not be able to detect this. The cost of our algorithm is broadly independent of the size of the state space. The Lotka-Volterra example illustrates the performance of our algorithm in a situation where the state space is unbounded and indicates that it can give a satisfactory performance in such situations. 
 
An interesting problem is the choice of $\lambda > \max_x Q(x)$. This is also an issue for Rao and Teh, but in their situation the answer is reasonably clear cut. For their algorithm, large values of $\lambda$ increase mobility and therefore efficiency, but at the same time increase cost, because the number of virtual jumps increases. The value of $\lambda$ is therefore the largest permitted by constraints of cost. 
 
In the situation here, the choice $\lambda = \max_x Q(x)$ can lead to unfortunately low acceptance probabilities; if (for example) point $x_i$ is proposed via $P(x_{i-1},.)$, but $x_i = x_{i+1}$, then for $\max_x Q(x) = Q(x_i)$, $P(x_i,x_{i+1}) = 0$, so that $a(\xx\p,\xx) = 0$. On the other hand, if $\lambda$ is too large, the skeleton will have more points, many of them virtual, which decreases efficiency. 

In conclusion, the algorithm presented here gives a new contribution, which complements existing approaches; some of the important applications are not within the scope of existing algorithms. 

The algorithm can also be extended in a straightforward manner to the situation where instead of the whole trajectory, the observed data is the trajectory sampled at a finite number of fixed time points.

\bibliographystyle{plainnat}
\bibliography{references}   

\end{document}